# A Fast and Memory-Efficient Algorithm for Searching Faint Moving Objects using Synthetic Tracking

Chengxing Zhai

Division of Emerging Interdisciplinary Areas, The Hong Kong University of Science and Technology, Clear Water Bay, Kowloon, Hong Kong

Albert J. Zhai

Department of Computer Science, University of Illinois, Urbana-Champaign, IL 61820, USA

Abstract

Optical imaging is a widely used method for object detection. While bright targets are easily detected using standard detect-before-track algorithms on single images, detecting faint moving objects with unknown trajectories remains challenging. Long exposure times are infeasible for such targets, as the resulting image streaking causes a loss of signal-to-noise ratio (S/N) due to energy dispersal across multiple pixels.

To address this, *synthetic tracking*—a track-before-detect (TBD) method—replaces long exposures with many short exposures, integrating them in post-processing to recover the moving signal. Enabled by modern CMOS detectors with fast readout and low read noise, this technique synthesizes telescope tracking by searching over a grid of unknown velocities. However, the conventional brute-force implementation, while functional, imposes a significant computational burden: integrating $N_f$ frames of size $(N_x, N_y)$ requires $O(N_x N_y N_f^3)$ operations. This cost becomes prohibitive for megapixel-scale frames with $N_f \sim 100$. Although dynamic programming can theoretically reduce the computational complexity by a factor of $N_f$, it introduces prohibitive memory requirements of $O(N_x N_y N_f^2)$, making it unsuitable for large-format detectors.

This article presents a novel algorithm that achieves the computational efficiency of dynamic programming while maintaining a memory footprint of just $O(N_x N_y N_f)$ by slightly approximating the computation of the number of pixels to shift for frame alignment. By expressing the shift as a sum of terms proportional to the digits of tracking velocity in binary and rounding up each of these terms to yield an integer shift, frames can be integrated efficiently by hierarchically adding intermediate sums that are shared across different velocities with the same higher bits in binary. The approximation only introduces an extra small tracking error (typically smaller than 0.7 pixel) that increases logarithmically with the number of frames. This advancement enables synthetic tracking to be executed with high energy efficiency, making it practical for deployment on standard workstations rather than requiring high-performance computing infrastructure such as GPU servers or CPU clusters.

## 1 Introduction

Track-before-detect (TBD) is an important paradigm for detecting faint targets with unknown or poorly constrained trajectories [1-2]. Detection difficulty increases substantially for rapidly moving sources, a scenario exemplified by small near-Earth objects (NEOs). These bodies are extremely faint at large geocentric distances but move rapidly once they approach Earth, at which point they may become marginally detectable. Discovering and monitoring small NEOs is a fundamental requirement for planetary-defense surveys for its large population [3]. Even a 30 m-diameter NEO can produce catastrophic damage upon impacting a densely populated region. We usually use long exposure to image faint targets to gain signal-to-noise ratio (S/N), however, motion-induced trailing loss arises when a moving target is imaged as an extended streak rather than a compact point-spread-function (PSF) spot [4]. Short-exposure imaging mitigates trailing loss; however, individual short-exposure frames do not contain sufficient signal energy to detect faint moving sources. Multiple short-exposure frames must therefore be co-added to boost the S/N. Based on these considerations, synthetic tracking has been developed to survey small NEOs [4-6]. Instead of acquiring a single long-exposure image, the technique operates on a 3-D image datacube composed of many short-exposure snapshots. In synthetic tracking, the integration of these short-exposure images numerically simulates on-sky tracking of a moving source by spatially shifting each frame according to a candidate target motion vector before the summation. Since the true target velocity is unknown, a grid of plausible velocities must be exhaustively searched. This brute-force search dominates the overall computational cost.

Although sub-pixel frame shifts can be implemented via interpolation, standard synthetic tracking pipelines only align the frames to integer-pixel precision, allowing a maximum tracking error of $\pm 0.5$ pixel [5]. For a datacube containing $N_f$ frames of dimension $(N_x, N_y)$, surveys are typically configured such that targets move no more than one PSF width, or approximately one pixel, between consecutive exposures [6]. Therefore, we usually consider the range of motion is within the range of 1 pixel per frame when integrating the frames. Velocity grids are therefore commonly sampled over the range $[0, 1, \cdots, N_f - 1]/N_f$. Neglecting directional degrees of freedom, the total computational cost scales as $N_x N_y N_f N_{v_x} N_{v_y}$ with the number of velocities $N_{v_x} \sim N_{v_y} \sim N_f$,with $v_x$ and $v_y$ being velocity components along $x$ and $y$, yielding an overall complexity of $O(N_x N_y N_f^3)$. For megapixel-scale frames with $N_f \sim 100$ frames, the total workload reaches the teraFLOP regime, necessitating massively parallel hardware. This computational bottleneck has motivated existing operational implementations, such as the JPL robotic telescope pipeline, to rely on GPU acceleration [6].

The fast discrete X-ray transform (FaXT) adapts a 2-D fast Radon transform algorithm built upon dynamic programming and divide-and-conquer principles [7] into a 3-D formulation, accelerating synthetic tracking by casting frame integration as a computed-tomography-style X-ray transform [8]. The computation of synthetic tracking has been significantly reduced to $\sim 2N_xN_yN_f^2$, $N_f/2$ times faster than the brute force approach. The primary practical limitation of FaXT is its substantial memory requirement of $N_xN_yN_f^2$.

We propose a novel frame integration strategy that matches the computational speed of dynamic programming-based methods while limiting memory consumption to $N_xN_yN_f$ comparable to the memory footprint of the raw input datacube, in sharp contrast to $N_xN_yN_f^2$ required by the FaXT algorithm. Our method makes an approximation in the computation of the integer shifts of frames for aligning them to simulate a tracking rate and then integrates the frames by adding consecutive frame pairs, which allows us to explicitly store the required intermediate frames for signal integration.

## 2 Method

We name the proposed algorithm Fast Synthetic Tracking (FST), which will be presented in this section.

### 2.1 Conventional Implementation

The brute force formulation of synthetic tracking can be summarized in the pseudocode below. We sample velocities $v_x$ and $v_y$ on a grid spanning approximately $\pm 1$ pixel/frame. It is convenient to consider a grid with spacing being $1/N_f$ so that the tracking rate error is $\pm 1/(2N_f)$ pixel/frame (maximum tracking error to be less than 0.5 pixel) and with candidate velocities taking discrete values $0, \pm 1/N_f, \pm 2/N_f, \cdots, \pm (N_f - 1)/N_f$. To simplify the discussion, let $N_f = 2^n$. Without loss of generality, let us consider the case where the candidate velocities are all positive. The cases with different signs can be done similarly by considering the frame shifts with the desired direction. Let $(x, y)$ denote the pixel coordinate in a frame, $I(x, y, f)$ represent the intensity of frame $f$, and $I_{ST}(x, y, v_x, v_y)$ be the integrated intensity with a linear tracking rate of $(v_x, v_y)$ respectively. We have the following pseudo-code (Python-like) for synthetic tracking integration as

detected_signals = []

for $v_x$ in $[0, 1, \cdots, N_f - 1]/N_f$:

  for $v_y$ in $[0, 1, \cdots, N_f - 1]/N_f$:

    $I_{ST}(x, y, v_x, v_y) = 0$

for $f$ in $[0, 1, \cdots, N_f - 1]$:

$I_{ST}(x, y, v_x, v_y) = I_{ST}(x, y, v_x, v_y) + I(x + \mathrm{round}(v_x f), y + \mathrm{round}(v_y f), f)$

signals_detected = $\mathrm{convolve}\left(I_{ST}(x, y, v_x, v_y), k(x, y)\right)$ > detection_threshold

detected_signals.append(signals_detected)

Here $k(x, y)$ denotes a spatial filter kernel used to improve the detection S/N. A matched filter, constructed from the instrument PSF, is generally optimal for optical imaging data [5]. The brute force implementation yields a computational complexity of $O(N_x N_y N_f^3)$. For typical parameters $N_x$ , $N_y \sim 10^3$ and $N_f \sim 10^2$, the workload falls within the teraFLOP range, requiring GPU hardware or large CPU clusters.

## 2.2 Fast Synthetic Tracking

Operational synthetic-tracking pipelines currently depend on GPU acceleration to manage computational cost [6]. Our FST algorithm introduces a controlled approximation to drastically reduce computation while maintaining a memory footprint of memory of $N_x N_y N_f$, matching the size of the input datacube.

The core mathematical approximation is to approximate the *round* function in $round(v_x f)$and $round(v_y f)$, for computing the integer shifts, as a sum of *round* of addends. To illustrate this, we use binary representations for $f$

$$f = \sum_{i=0}^{n-1} b_f^i \, 2^i,$$

to have

$$round(v_x f) = round\left(\sum_{m=0}^{n-1} v_x \; b_f^m \, 2^m\right) \approx \sum_{m=0}^{n-1} round\left(v_x b_f^m 2^m\right) = \sum_{m=0}^{n-1} b_f^m \, dx_m$$

where we have defined

$$dx_m \equiv round(v_x 2^m).$$

Using the binary representations for $v_x$,

$$v_x = \sum_{i=0}^{n-1} b_{v_x}^i 2^{i-n},$$

with $b_{v_x}^i, i = 1,2, \cdots, n-1$ being bits, we compute

$$dx_m = round(v_x 2^m) = b_{v_x}^{n-m-1} + \sum_{i=n-m}^{n-1} b_{v_x}^{i} 2^{i+m-n}$$

Similarly for y direction, defining $dy_m \equiv round(v_y 2^m)$, we have

$$round(v_y f) \approx \sum_{m=0}^{n-1} b_f^m \, dy_m \ , dy_m = round(v_y 2^m) = b_{v_y}^{n-m-1} + \sum_{i=n-m}^{n-1} b_{v_y}^{i} 2^{i+m-n}$$

with $b_{v_y}^i$ presenting bits for $v_y$ in binary. The integration of frames or summing over the frame index $f$ can be viewed as summing over its bits $b_f^m, m = 0,1, \cdots n-1$ in its binary representation. With the round function approximation, the sum over $b_f^m$'s can be carried out independently because frame shifts are decoupled between the bits $b_f^m$'s.

Summing over bits $b_f^m$'s can be conveniently viewed as first summing over consecutive paired frames (lowest bit) and then summing over the consecutive resulting frames (the second lowest bits) and so on. This recursive formulation for frame integration is illustrated in Fig. 1 and mathematically described below.

Define the initial 3-D datacube

$$I^{(0)}(x, y, b_f^0, b_f^1, \cdots, b_f^{n-1}) \equiv I(x, y, f),$$

where the frame index has been explicitly expressed in terms of bits $b_f^i$'s. The standard synthetic-tracking summation

$$I(x, y, v_x, v_y) = \sum_{f=0}^{N_f - 1} I(x + round(v_x f), y + round(\ v_y f), f)$$

is rewritten using our bitwise decomposition and rounding approximation:

$$\begin{aligned} I(x, y, v_x, v_y) &= \prod_{i=0}^{n-1} \sum_{b_f^i = 0}^{1} I^{(0)}(x + round(v_x f), y + round(\ v_y f), b_f^0, b_f^1, \cdots, b_f^{n-1}) \\ &\approx \prod_{i=0}^{n-1} \sum_{b_f^i = 0}^{1} I^{(0)}\left(x + \sum_{m=0}^{n-1} b_f^m \, dx_m , y + \sum_{m=0}^{n-1} b_f^m \, dy_m , b_f^0, b_f^1, \cdots, b_f^{n-1}\right). \end{aligned}$$

We perform recursion by summing sequentially over each binary bit. After summing over $b_f^0$, we obtain

$$I(x, y, v_x, v_y) = \prod_{i=1}^{n-1} \sum_{b_f^i=0}^{1} I^{(1)}\left(x + \sum_{m=1}^{n-1} b_f^m \, dx_m \, , y + \sum_{m=1}^{n-1} b_f^m \, dy_m \, , b_f^1, \cdots, b_f^{n-1}\right)$$

where we have introduced

$$I^{(1)}(x, y, b_f^1, \cdots, b_f^{n-1}) \equiv I^{(0)}(x, y, 0, b_f^1, \cdots, b_f^{n-1}) + I^{(0)}(x + dx_0, y + dy_0, 1, b_f^1, \cdots, b_f^{n-1})$$

with the base-level shift $(dx_0, dy_0) = round(v_x, v_y) = (b_{v_x}^{n-1}, b_{v_y}^{n-1})$.

The general recursive update rule for stage m reads:

$$I(x, y, v_x, v_y) = \prod_{i=m}^{n-1} \sum_{b_f^i=0}^{1} I^{(m)}\left(x + \sum_{m\prime=m}^{n-1} b_f^{m\prime} dx_{m\prime} \, , y + \sum_{m\prime=m}^{n-1} b_f^{m\prime} \, dy_{m\prime} \, , b_f^m, \cdots, b_f^{n-1}\right)$$

where

$$I^{(m)}(x, y, b_f^m, \cdots, b_f^{n-1}) \\ \equiv I^{(m-1)}(x, y, 0, b_f^m, \cdots, b_f^{n-1}) + I^{(m-1)}(x+dx_{m-1}, y+dy_{m-1}, 1, b_f^m, \cdots, b_f^{n-1}).$$

The recursion terminates at $m = n$ yielding the final integrated image $I(x, y, v_x, v_y) = I^{(n)}(x, y)$.

At recursion level m, there are, there are totally $2^{n-m}$ intermediate arrays $I^{(m+1)}(x, y, b_f^m, \cdots, b_f^{n-1})$ in view of the all the possible values of bits $b_f^m, \cdots, b_f^{n-1}$.

As sketched in Figure 1, the peak memory consumption scales as the total number of frames $\sum_{m=0}^{n} 2^m = 2^{n+1} - 1$ giving the amount of $\sim 2^{n+1} N_x N_y = 2 N_x N_y N_f$.

Because the computation of $I^{(m)}(x, y, b_f^m, \cdots, b_f^{n-1})$ only needs $(dx_i, dy_i), i = 1, \cdots, m-1$ and $(dx_i, dy_i)$ only depends on the $(i+1)$th highest bits of velocity vector $\left(b_{v_x}^{n-j-1}, b_{v_y}^{n-j-1}, j = 0, 1, \cdots i\right)$, we therefore can arrange the computation with nested loops over the velocity vector bits. Starting from the highest bit and compute $I^{(m)}(x, y, b_f^m, \cdots, b_f^{n-1})$ at the outer loops whenever possible to minimize the number of computations. Therefore, $I^{(m)}(x, y, b_f^m, \cdots, b_f^{n-1})$ can be computed at m-level of the loops. The pseudocode for the full recursive algorithm is:

for $b_{v_x}^{n-1}$ $in$ $[0, 1]$:

  for $b_{v_y}^{n-1}$ $in$ $[0, 1]$:

$$I^{(1)}(x, y, b_f^1, \cdots, b_f^{n-1}) = I^{(0)}(x, y, 0, b_f^1, \cdots, b_f^{n-1}) + I^{(0)}\left(x + b_{v_x}^{n-1}, y + b_{v_y}^{n-1}, 1, b_f^1, \cdots, b_f^{n-1}\right)$$

⋮

for $b_{v_x}^{n-m}$ $in$ $[0,1]$:

for $b_{v_y}^{n-m}$ $in$ $[0,1]$:

$$I^{(m)}\left(x,y,b_f^1,\cdots,b_f^{n-1}\right) = I^{(m-1)}\left(x,y,0,b_f^m,\cdots,b_f^{n-1}\right) + I^{(m-1)}\left(x+dx_m,y+dy_m,1,b_f^m,\cdots,b_f^{n-1}\right)$$

for $b_{v_x}^0$ $in$ $[0,1]$ :

for $b_{v_y}^0$ $in$ $[0,1]$:

$$I^{(n)}(x,y) = I^{(n-1)}(x,y,0) + I^{(n-1)}(x+dx_n,y+dy_n,1)$$

Note that each loop involving velocity vector along two dimensions, therefore, at level-m the computation amount has a factor $4^m$ from the nested loops, and we need to add $2^{n-m}$ frame-pairs (different $b_f^i$'s), therefore summing the arithmetic workload, the total computational complexity is

$$N_x N_y \sum_{m=1}^{n} 4^m \times 2^{n-m} = N_x N_y 2^n (2^{n+1}-1) = N_x N_y N_f (2N_f - 1).$$

The integrated frame can be passed into a detection threshold checking logic to create a list of detected signals.

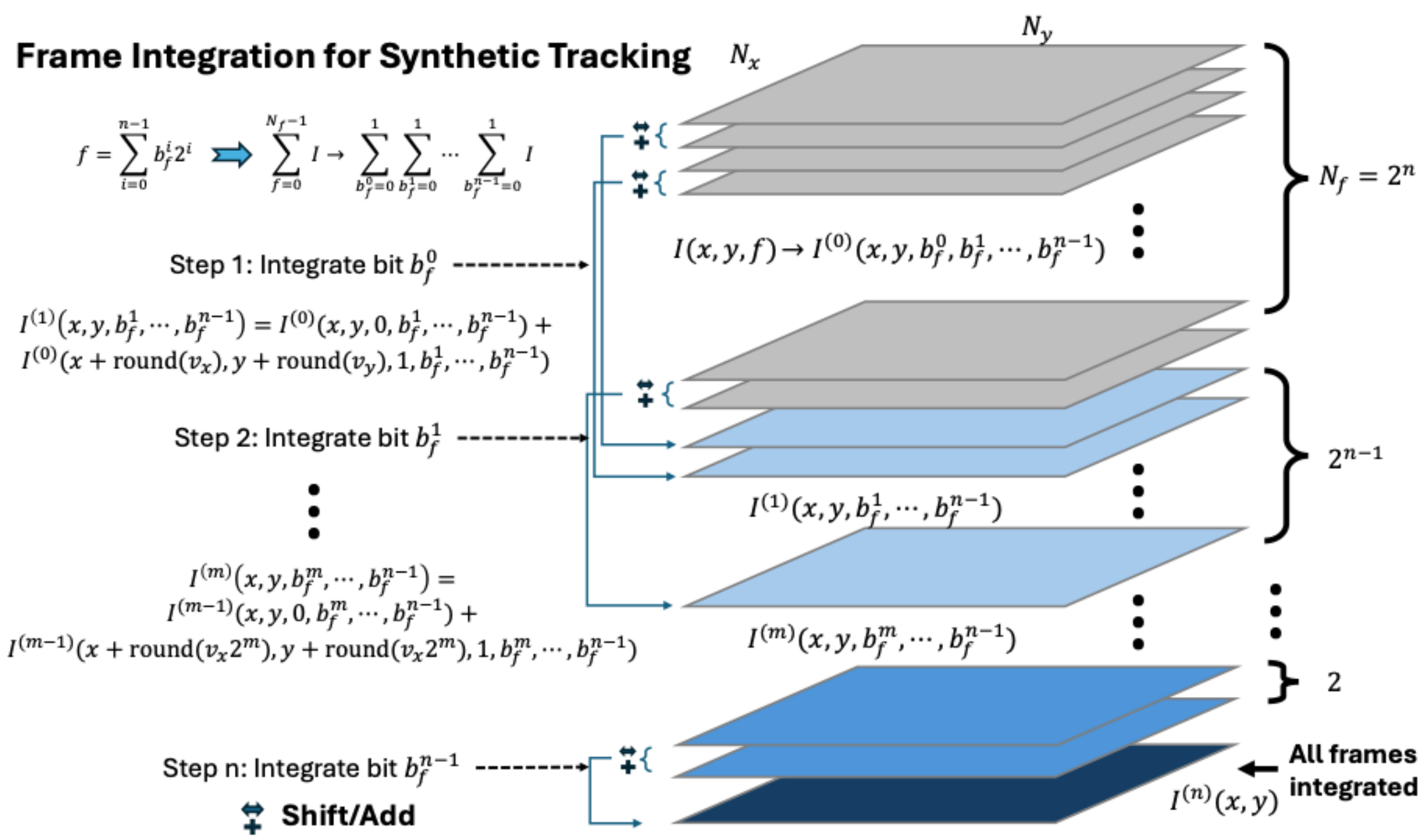

**Figure 1**. A schematic chart illustrating how the frames are integrated by pairing consecutive frames and the layout of memory usage to store the intermediate data frames.

So far, we have assumed the velocity range is up to 1 pixel/frame. In case the range is larger with x, y velocity components ranges being up to $V_x$ pixel/frame and $V_y$ pixel/frame, the total computation load is then $N_x N_y N_f (2N_f - 1) V_x V_y \sim 2 N_x N_y N_{v_x} N_{v_y}$, where we have used $N_{v_x} = V_x N_f, N_{v_y} = V_y N_f$. Compared with the brute force load of $N_x N_y N_{v_x} N_{v_y} N_f$, the speed up is a factor of $N_f/2$, which is significant for $N_f \sim 100$.

## 3 Results

In this subsection, we present results from applying FST to both simulation data and observational data from the Zwicky Transient Facility (ZTF).

### 3.1 Simulation Tests

We present single-thread CPU execution times of FST measured on an Apple Macbook Pro M3 Max processor for varying spatial dimensions and number of frames.

**Table 1.** Single-thread CPU runtime for synthetic-tracking integration of 64 frames on an Apple M3-Max processor

| CPU Time (sec) | $N_x$ | $N_y$ | $N_f$ |
|---|---|---|---|
| 620 | 1000 | 1000 | 64 |
| 1230 | 1000 | 2000 | 64 |
| 1231 | 2000 | 1000 | 64 |
| 2448 | 2000 | 2000 | 64 |

Measured runtime scales linearly with $N_x$ and $N_y$, consistent with our complexity analysis. The execution time also scales quadratically with number of frames $N_f$ as plotted in Figure 2. Note that the quoted timings here cover the full two-dimensional velocity search space (positive and negative velocity components); processing only the positive-velocity quadrant would reduce runtime by approximately a factor of four as considered in the estimation presented in section 2.2.

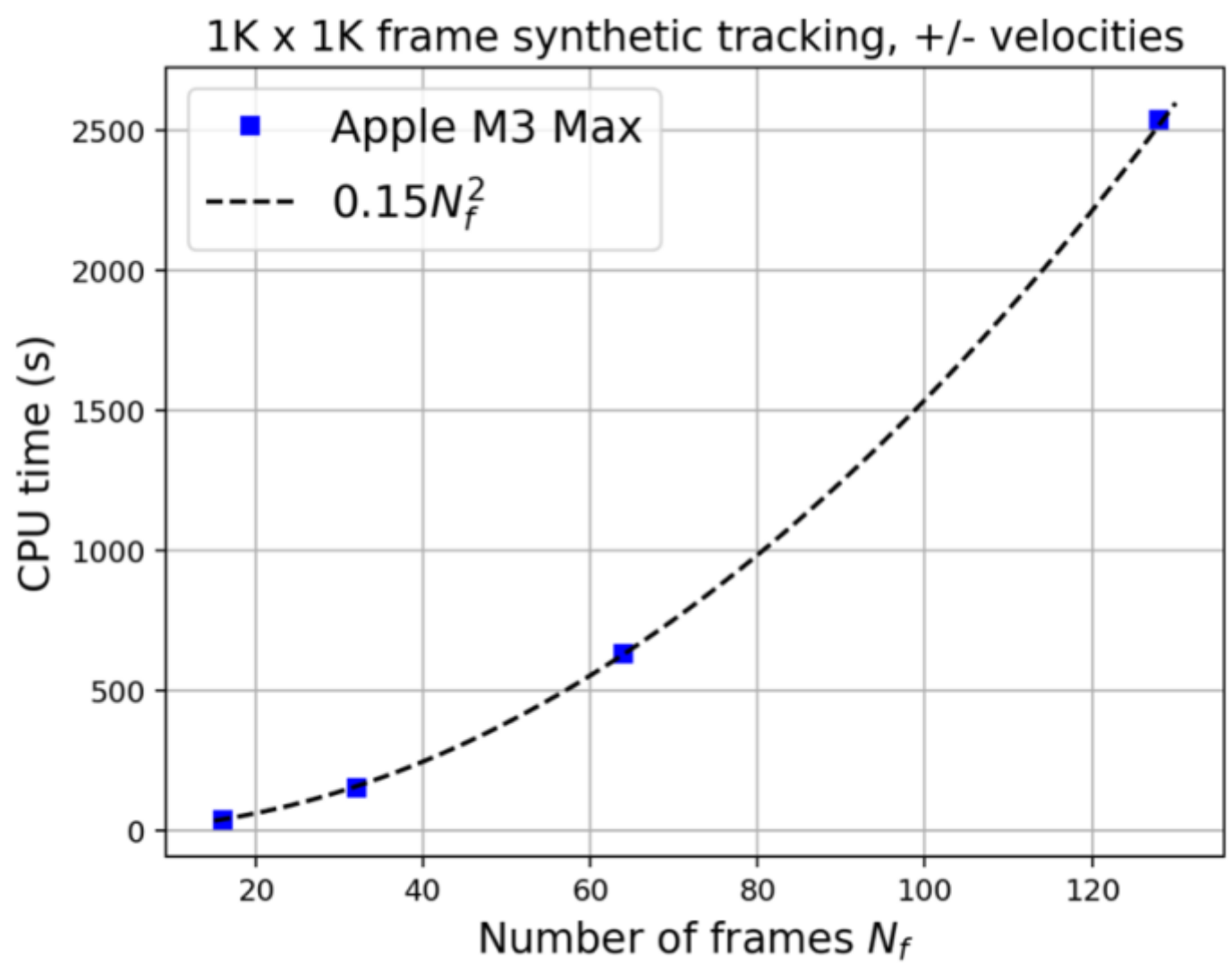

**Figure 2.** CPU time measured on Apple Max-M3 processor for using FST to integrate different number of frames of size (1K*1K).

FST approximates the round of a sum as the sum of round values. This approximation introduces extra errors from the additional round errors, whose number is $\sim \log_2 N_f$ with random values within range [-0.5, 0.5]. This is the price paid to use FST compared with the brute force approach discussed in section 2.1.

We numerically evaluated the tracking errors for integrating different number of frames $N_f$ up to 1024. The true velocities are simulated with a quantum of $2^{-13}$ to cover range of [0,1]. The maximum tracking errors is plotted against number of frames integrated shown in the left plot in Figure 3. It has an upper bound about $\sim \frac{1}{6}\log_2 N_f + 0.5$. An upper bound of $\frac{1}{6}\log_2 N_f$ was reported by reference [7] where the definition of tracking error has a minor difference due to our distinct geometric formulation. A detailed explanation is provided in the Appendix.

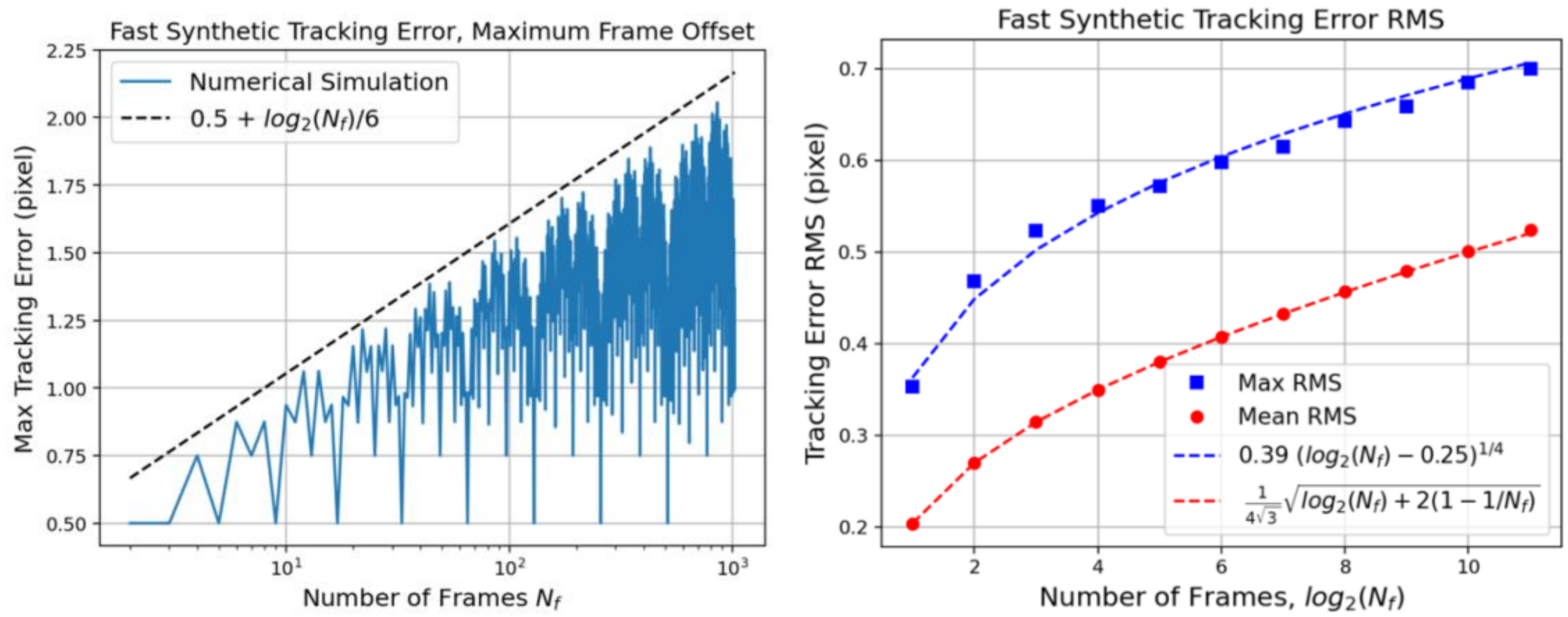

**Figure 3.** Tracking errors of FST from simulation as function of number of frames: left plot displays maximum tracking error and right plot displays the mean of tracking error RMS (root-mean-squares) for different velocities.

The right plot in Figure 3 displays the mean and maximum of RMS (root-mean-squares) of the tracking error over the integrated frames. This quantifies how much larger the effective PSF is due to the tracking errors. As expected, it increases with the number of frames logarithmically. We also show in the Appendix the behavior of mean RMS is $\frac{1}{4\sqrt{3}}\sqrt{\log_2 N_f + 2(1 - N_f^{-1})}$. The price paid for speeding up the computation to integrate 1000 frames is less than 0.7 pixel. Unless the PSF is smaller than 1 pixel, this extra error is in general small.

Because of the extra spread of the integrated signal from these tracking errors, the detection S/N is slightly lower than from the synthetic tracking with only the integer pixel shift error (+/-0.5 pixel). The tracking error would degrade the detection S/N relative the conventional implementation which only has tracking error up to $\pm 0.5$ pixel. With the quantification of the tracking error RMS, we can estimate the effective PSF after using FST. We did a simulation for different sizes of PSFs assuming a Gaussian shape. We then estimate the optimal kernel $k(x, y)$ to be used to perform matched filter detection. The SNR degradation due FST relative to conventional implementation is displayed in Figure 4 as function of the size of PSF, which is measured in unites of Gaussian $1\sigma$.

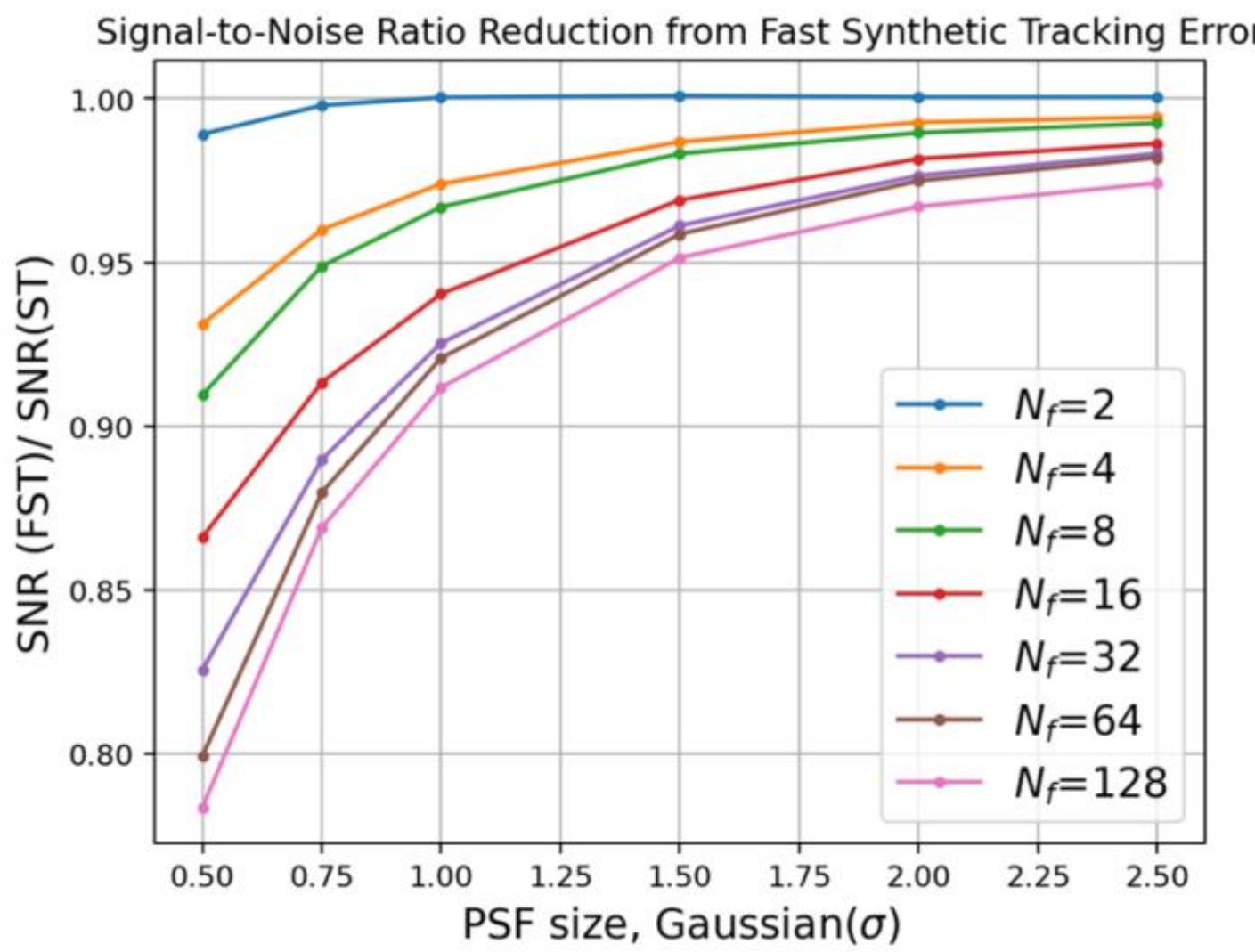

**Figure 4.** Degradation of signal-to-noise ratio (SNR) as function of PSF size.

Figure 5 displays examples of the integrated signal from integrating 128 frames for simulated six objects with different motion rates and PSF sizes.

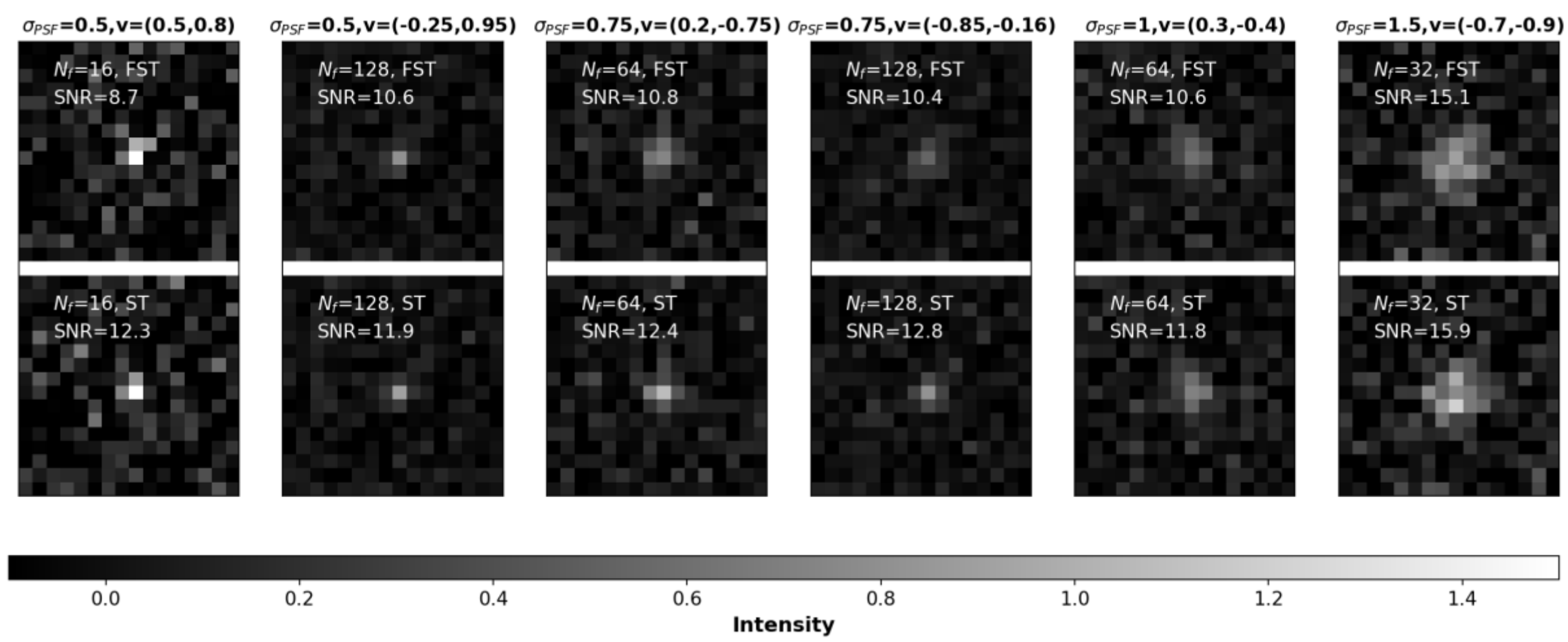

**Figure 5**. Example PSFs after FST and conventional synthetic tracking of six objects used in simulation.

## 3.2 Application to Zwicky Transit Facility Datasets

As an application to detecting asteroids, we applied FST to integrate a deep drilling dataset from the Zwicky Transient Facility (ZTF) as demonstrated by reference [9]. Figure 6 displays what we discovered from integrating a datacube, taken on 2018 July 12, for which we integrated 128 frames over the central 2000x2000 pixel area. The rates of motion suggest that these objects are likely main belt asteroids except for the first one, which is probably a Trojan asteroid going with the Jupiter. The number of detections is roughly 8 over about 0.3

Sqdeg. For ZTF's large FOV of ~ 47 Sqdeg, we expect to detect about 1 thousand objects, which is consistent with the statistics reported by reference [9].

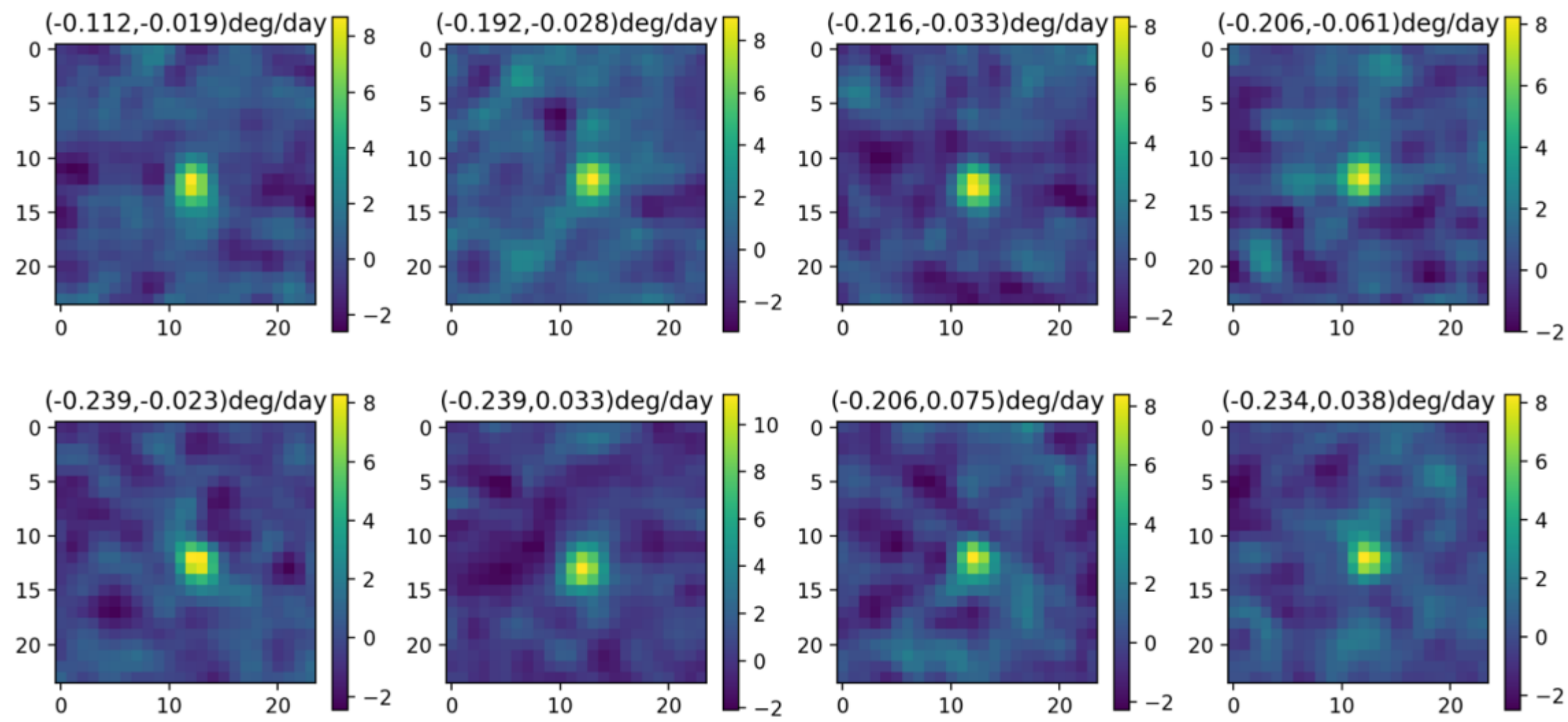

**Figure 6**. Example detections from integration 128 frames of a ZTF dataset over the center 2000x2000 pixels. The rates suggest that they are main-belt asteroids except for the first one, which is likely to be a Trojan asteroid moving with the Jupyter's orbit.

## 4 Conclusions and Discussions

We have introduced the Fast Synthetic Tracking (FST) algorithm for detecting faint moving astronomical targets from multi-frame image datacubes. Our implementation achieves the $O(N_x N_y N_f^2)$ computational speed-up of dynamic-programming-based FaXT yet limits memory consumption to $O(N_x N_y N_f)$, comparable to the raw input data volume, whereas FaXT requires $O(N_x N_y N_f^2)$ memory. This algorithm enables synthetic-tracking processing on consumer-grade workstations and laptops, removing the historical requirement for dedicated GPU servers or large CPU clusters. It therefore lowers the deployment barrier for NEO search pipelines supporting planetary-defense efforts. FST only introduces a small amount of tracking errors (less than 0.7 pixel RMS for integrating 1000 frames), growing logarithmically with number of frames.

Beyond asteroid surveys, the algorithm can be adapted to one-dimensional streak-detection tasks, where it again offers substantial memory savings relative to conventional fast Radon-transform methods [7,10]. Straightforward extensions also permit processing of four-dimensional datacubes. We anticipate that this low-memory fast formulation will facilitate broader adoption of synthetic-tracking techniques across additional application domains.

The code that we implemented the FST together with a test routine is available at *https://github.com/ajzhai/fast-synthetic-tracking*.

## Acknowledgements

The authors would like to acknowledge the usage of observations obtained with the Samuel Oschin 48-inch Telescope at the Palomar Observatory as of the Zwicky Transient Facility project. ZTF is supported by the National Science Foundation under Grant No. AST-1440341 and a collaboration including Caltech, IPAC, the Weizmann Institute for Science, the Oskar Klein Center at Stockholm University, the University of Maryland, the University of Washington, Deutsches Elektronen Synchrotron and Humboldt University, Los Alamos National Laboratories, the TANGO Consortium of Taiwan, the University of Wisconsin at Milwaukee, and Lawrence Berkeley National Laboratories. Operations are conducted by COO, IPAC, and UW.

## Appendix

Here we explain the upper bound of the behavior of maximum tracking error $\sim 1/6 \log_2 N_f$.

Let $v_x = \sum_{i=-\infty}^{n-1} b_{v_x}^i 2^{m+i-n}$. FST tracking error can be computed as

$$\Delta x = v_x f - \sum_{m=0}^{n-1} b_f^m \, dx_m$$

$$= \sum_{i=-\infty}^{n-1} \sum_{m=0}^{n-1} b_{v_x}^i b_f^m 2^{m+i-n} - \sum_{m=0}^{n-1} b_f^m \left( b_{v_x}^{n-m-1} + \sum_{i=n-m}^{n-1} b_{v_x}^i 2^{i+m-n} \right)$$

$$= \sum_{m=0}^{n-1} b_f^m \left( \sum_{i=-\infty}^{n-m-2} b_{v_x}^i 2^{m+i-n} - \frac{1}{2} b_{v_x}^{n-m-1} \right) = \sum_{m=0}^{n-1} b_f^m \left( \sum_{i=0}^{\infty} b_{v_x}^{n-m-2-i} 2^{-2-i} - \frac{1}{2} b_{v_x}^{n-m-1} \right)$$

If we want to maximize the round error of *round*($2^m v_x$), the bit pattern of $v_x$ cannot be uniformly 1 or 0 because the round error would be 0. The maximum round error can be achieved when $v_x = 2^{-m-1}$ or $b_{v_x}^i = \delta_{i,n-m-1}$. But the total error is no more than $2^{-1} + 2^{-2} + 2^{-3} + \cdots < 1$. Assuming a periodic pattern, we found that having a bit pattern of 0, 1, $0, 1, 0, \cdots$ would maximize the total errors. We can compute the round errors to be 1/2 -1/8 - 1/32 $+ \cdots \sim$ 1/2 -1/8/(1-1/4) =1/3. In the meantime, we have $n = \log_2 N_f$ bits, so we alternatively taking the same sign of round errors, we also alternatively having bits to be 1's for $b_{v_x}^i$, so that each 1 would match a round error of 1/3, thus ½ $\log_2 N_f$ bits to be 1's alternatively with each 1 gives 1/3 error, thus the asymptotic behavior of 1/6 $\log_2 N_f$.

We now look at the average RMS. It is convenient to consider shifted bits:

$$\tilde{b}_f^m \equiv b_f^m - \frac{1}{2}, \tilde{b}_{v_x}^i \equiv b_{v_x}^i - \frac{1}{2}$$

so that these bits fluctuate around 0 with values of $\pm 1/2$. The tracking error can be expressed as

$$\Delta x = \sum_{m=0}^{n-1} \tilde{b}_f^m \left( \sum_{i=0}^{\infty} b_{v_x}^{n-m-2-i} 2^{-2-i} - \frac{1}{2} b_{v_x}^{n-m-1} \right) + \frac{1}{2} \sum_{m=0}^{n-1} \left( \sum_{i=0}^{\infty} b_{v_x}^{n-m-2-i} 2^{-2-i} - \frac{1}{2} b_{v_x}^{n-m-1} \right).$$

Because the errors from rounding numbers in general are unbiased, so the second term, which is the sum of these errors tend to be small. Now we show it is in range of [-0.5, 0.5]. First rearranging the terms gives

$$\frac{1}{2}\sum_{m=0}^{n-1}\left(\sum_{i=0}^{\infty} b_{v_x}^{n-m-2-i}2^{-2-i}-\frac{1}{2}b_{v_x}^{n-m-1}\right)=\frac{1}{4}\sum_{m=0}^{n-1}\left(\sum_{i=0}^{\infty} b_{v_x}^{m-1-i}2^{-1-i}-b_{v_x}^{m}\right)$$

$$=\frac{1}{4}\left(-\sum_{m=0}^{n-1} b_{v_x}^{m}+\sum_{i=0}^{\infty}\sum_{m=0}^{n-1} b_{v_x}^{m-1-i}2^{-1-i}\right)=\frac{1}{4}\left(-\sum_{m=0}^{n-1} b_{v_x}^{m}+\sum_{i=1}^{\infty}\sum_{m=0}^{n-1} b_{v_x}^{m-i}2^{-i}\right).$$

For the second term, shifting index $m \to m-i$ and switching the order of sum, we have

$$\frac{1}{2}\sum_{m=0}^{n-1}\left(\sum_{i=0}^{\infty} b_{v_x}^{n-m-2-i}2^{-2-i}-\frac{1}{2}b_{v_x}^{n-m-1}\right)=\frac{1}{4}\left(-\sum_{m=0}^{n-1} b_{v_x}^{m}+\sum_{i=1}^{\infty}\sum_{m=-i}^{n-1-i} b_{v_x}^{m}2^{-i}\right)$$

$$=\frac{1}{4}\left(-\sum_{m=0}^{n-1} b_{v_x}^{m}+\sum_{m=0}^{n-2}\sum_{i=1}^{n-1-m} b_{v_x}^{m}2^{-i}+\sum_{m=-\infty}^{-1}\sum_{i=-m}^{n-1-m} b_{v_x}^{m}2^{-i}\right)$$

$$=\frac{1}{4}\left(-\sum_{m=0}^{n-1} b_{v_x}^{m}+\sum_{m=0}^{n-2} b_{v_x}^{m}\sum_{i=1}^{n-1-m} 2^{-i}+\sum_{m=-\infty}^{-1} b_{v_x}^{m}\sum_{i=-m}^{n-1-m} 2^{-i}\right)$$

$$=\frac{1}{4}\left(-\sum_{m=0}^{n-1} b_{v_x}^{m}+\sum_{m=0}^{n-2} b_{v_x}^{m}(1-2^{-n+1+m})+\sum_{m=-\infty}^{-1} b_{v_x}^{m}(2^{m+1}-2^{-n+1+m})\right)$$

$$=\frac{1}{4}\left(-b_{v_x}^{n-1}-\sum_{m=0}^{n-2} b_{v_x}^{m}2^{-n+1+m}+(1-2^{-n})\sum_{m=-\infty}^{-1} b_{v_x}^{m}2^{m+1}\right)$$

where we can bound the negative part in the parenthesis by -2 when all the bits are one by summing the geometric series and the positive part in the parenthesis by 2, thus the total value can only be within $\pm 0.5$.

Expanding the square and using $\langle \tilde{b}_f^m \tilde{b}_f^n\rangle=\frac{1}{4}\delta_{m,n}$, $\langle \tilde{b}_f^m\rangle=0$, we compute the RMS of tracking error $\Delta x$ as

$$\langle(\Delta x)^2\rangle=\left\langle\left[\sum_{m=0}^{n-1}\tilde{b}_f^m\left(\sum_{i=0}^{\infty} b_{v_x}^{n-m-2-i}2^{-2-i}-\frac{1}{2}b_{v_x}^{n-m-1}\right)\right]^2\right\rangle+\frac{1}{4}\left\langle\left[\sum_{m=0}^{n-1}\left(\sum_{i=0}^{\infty} b_{v_x}^{n-m-2-i}2^{-2-i}-\frac{1}{2}b_{v_x}^{n-m-1}\right)\right]^2\right\rangle$$

$$=\frac{1}{4}\sum_{m=0}^{n-1}\left\langle\left(\sum_{i=0}^{\infty} b_{v_x}^{n-m-2-i}2^{-2-i}-\frac{1}{2}b_{v_x}^{n-m-1}\right)^2\right\rangle+\frac{1}{4}\left\langle\left[\sum_{m=0}^{n-1}\left(\sum_{i=0}^{\infty} b_{v_x}^{n-m-2-i}2^{-2-i}-\frac{1}{2}b_{v_x}^{n-m-1}\right)\right]^2\right\rangle$$

Because $\sum_{i=0}^{\infty} 2^{-2-i}=1/2$, the overall shift of ½ cancels between the two terms and we get

$$\sum_{i=0}^{\infty} b_{v_x}^{n-m-2-i}2^{-2-i}-\frac{1}{2}b_{v_x}^{n-m-1}=\sum_{i=0}^{\infty} \tilde{b}_{v_x}^{n-m-2-i}2^{-2-i}-\frac{1}{2}\tilde{b}_{v_x}^{n-m-1}$$

and

$$\langle(\Delta x)^2\rangle = \frac{1}{4}\sum_{m=0}^{n-1}\left\langle\left(\sum_{i=0}^{\infty}\tilde{b}_{v_x}^{n-m-2-i}2^{-2-i} - \frac{1}{2}\tilde{b}_{v_x}^{n-m-1}\right)^2\right\rangle + \frac{1}{4}\left\langle\left[\sum_{m=0}^{n-1}\left(\sum_{i=0}^{\infty}\tilde{b}_{v_x}^{n-m-2-i}2^{-2-i} - \frac{1}{2}\tilde{b}_{v_x}^{n-m-1}\right)\right]^2\right\rangle$$

Using $\langle\tilde{b}_{v_x}^m\tilde{b}_{v_x}^n\rangle = \frac{1}{4}\delta_{m,n}$ and expanding the square, we calculate the first term as

$$\frac{1}{4}\sum_{m=0}^{n-1}\langle\left(\sum_{i=0}^{\infty}b_{v_x}^{n-m-2-i}2^{-2-i} - \frac{1}{2}b_{v_x}^{n-m-1}\right)^2\rangle$$

$$= \frac{1}{4}\frac{1}{4}\sum_{m=0}^{n-1}\frac{1}{4}\left(1 + \frac{1}{4} + \frac{1}{16} + \cdots\right) = \frac{1}{16}n\frac{1}{4}\frac{1}{1-1/4} = \frac{1}{48}n$$

We evaluate the second term by noting that

$$\sum_{m=0}^{n-1}\left(\sum_{i=0}^{\infty}\tilde{b}_{v_x}^{n-m-2-i}2^{-2-i} - \frac{1}{2}\tilde{b}_{v_x}^{n-m-1}\right) = \frac{1}{2}\left[-\tilde{b}_{v_x}^{n-1} - \sum_{m=0}^{n-2}\tilde{b}_{v_x}^{m}2^{-n+1+m} + (1-2^{-n})\sum_{m=-\infty}^{-1}\tilde{b}_{v_x}^{m}2^{m+1}\right]$$

and using $\langle\tilde{b}_{v_x}^m\tilde{b}_{v_x}^n\rangle = \frac{1}{4}\delta_{m,n}$ to get

$$\left\langle\left[\sum_{m=0}^{n-1}\left(\sum_{i=0}^{\infty}\tilde{b}_{v_x}^{n-m-2-i}2^{-2-i} - \frac{1}{2}\tilde{b}_{v_x}^{n-m-1}\right)\right]^2\right\rangle = \frac{1}{4}\left[\frac{1}{4} + \sum_{m=0}^{n-2}2^{-2n+2m} + (1-2^{-n})^2\sum_{m=-\infty}^{-1}2^{2m}\right]$$

$$= \frac{1}{4}\left[\frac{1}{4} + 2^{-2n}\sum_{m=0}^{n-2}2^{2m} + \frac{1}{4}\frac{(1-2^{-n})^2}{1-\frac{1}{4}}\right] = \frac{1}{16} + \frac{1}{48}\left(1 - 2^{2-2n} + 4(1-2^{-n})^2\right)$$

$$= \frac{1}{6}(1-2^{-n}).$$

Therefore,

$$\langle(\Delta x)^2\rangle = \frac{n}{48} + \frac{1}{24}(1-2^{-n})$$

the RMS of tracking error $\sqrt{\langle(\Delta x)^2\rangle} = \frac{1}{4\sqrt{3}}\sqrt{\log_2 N_f + 2\left(1 - \frac{1}{N_f}\right)}$.